\documentclass[conference,letterpaper, 10pt]{IEEEtran}
\IEEEoverridecommandlockouts

\usepackage[utf8]{inputenc}
\usepackage[T1]{fontenc}
\usepackage[american]{babel}
\usepackage[center]{caption}
\usepackage{graphicx}
\usepackage{siunitx}
\usepackage{longtable,tabularx}
\usepackage{caption}
\usepackage{subfigure}
\usepackage{subcaption}
\usepackage{listings}
\usepackage{float}
\usepackage{amsmath,amssymb,exscale}
\usepackage{blindtext, graphicx}
\usepackage{verbatim}
\usepackage{algorithm}
\usepackage{algpseudocode}
\usepackage{fancyvrb}
\usepackage{bera}
\usepackage{amsmath}
\usepackage{amsthm}
\usepackage{amsfonts}

\usepackage{mathtools}
\usepackage{lipsum}
\usepackage[dvipsnames]{xcolor}
\usepackage{cuted}
\usepackage{flushend}
\usepackage{dblfloatfix} 
\usepackage{times}
\usepackage{placeins}

\usepackage{tikz}
\usetikzlibrary{automata, positioning, arrows, calc}

\tikzstyle{block} = [draw, fill=white, rectangle, minimum height=3em, minimum width=4em, thick]
\tikzstyle{sum} = [draw, fill=white, circle, thick]
\tikzstyle{input} = [coordinate]
\tikzstyle{noise} = [coordinate]
\tikzstyle{output} = [coordinate]
\tikzstyle{output1} = [coordinate]
\tikzstyle{disturbance} = [coordinate]
\tikzstyle{pinstyle} = [pin edge={to-,thin,black}]

\usepackage{scalerel}

\usepackage[bookmarks=false, hidelinks]{hyperref}
\usepackage{cleveref}
\crefname{equation}{Eq.}{Eqs.} 

\crefname{figure}{Fig.}{Figs.}

\def\BibTeX{{\rm B\kern-.05em{\sc i\kern-.025em b}\kern-.08em
T\kern-.1667em\lower.7ex\hbox{E}\kern-.125emX}}

\makeatletter 
\let\old@ps@headings\ps@headings 
\let\old@ps@IEEEtitlepagestyle\ps@IEEEtitlepagestyle 
\def\confheader#1{%
\def\ps@headings{%
\old@ps@headings%
\def\@oddhead{\strut\hfill#1\hfill\strut}%
\def\@evenhead{\strut\hfill#1\hfill\strut}%
}%
\def\ps@IEEEtitlepagestyle{%
\old@ps@IEEEtitlepagestyle%
\def\@oddhead{\strut\hfill#1\hfill\strut}%
\def\@evenhead{\strut\hfill#1\hfill\strut}%
}%
\ps@headings%
} 
\makeatother 

\usepackage[letterpaper, margin=.75in]{geometry}
\newgeometry{top=1in, bottom=0.75in, left=0.75in, right=0.75in}

\begin{document}
\title{Robust Attitude Control of Nonlinear UAV Dynamics with LFT Models and $\mathcal{H}_\infty$ Performance} 


\author{\IEEEauthorblockN{Tanay Kumar\thanks{Graduate Student, \texttt{ktanay@tamu.edu}} \hspace{0.5in} Raktim Bhattacharya \thanks{Professor, \texttt{raktim@tamu.edu}}\vspace{.1in} }
\IEEEauthorblockA{Aerospace Engineering, Texas A\&M University,\\ College Station, TX, 77843-3141.
}}

\maketitle
\begin{abstract}
Attitude stabilization of unmanned aerial vehicles in uncertain environments presents significant challenges due to nonlinear dynamics, parameter variations, and sensor limitations. This paper presents a comparative study of $\mathcal{H}_\infty$ and classical PID controllers for multi-rotor attitude regulation in the presence of wind disturbances and gyroscope noise. The flight dynamics are modeled using a linear parameter-varying (LPV) framework, where nonlinearities and parameter variations are systematically represented as structured uncertainties within a linear fractional transformation formulation. A robust controller based on $\mathcal{H}_\infty$ formulation is designed using only gyroscope measurements to ensure guaranteed performance bounds. Nonlinear simulation results demonstrate the effectiveness of the robust controllers compared to classical PID control, showing significant improvement in attitude regulation under severe wind disturbances.
\end{abstract}

\begin{IEEEkeywords}
LFT Modeling, LPV Systems, $\mathcal{H}_\infty$ Optimal Control, Flight Control.
\end{IEEEkeywords}

\section{Introduction}
Unmanned Aerial Vehicles (UAVs) have seen rapid adoption across diverse sectors, driving increased demands for robust control systems. Complex missions involving urban navigation, multi-agent coordination, and beyond-visual-line-of-sight operations present challenges that conventional control approaches struggle to address. These applications face significant disturbances, including atmospheric turbulence, wind gradients, mass distribution variations, sensor imprecision, and model uncertainties. Robust control methodologies are therefore essential to maintain stability and performance under such perturbations. Recent statistical analyses confirm this need, showing increased incidents related to control system failures in UAV operations \cite{grindley2024over}.

Linear controllers remain predominant in practical implementations due to their analytical tractability and computational efficiency. Proportional-Integral-Derivative (PID) control structures, particularly in cascaded configurations, are extensively deployed in industry \cite{lopez2023pid}; however, their single-input-single-output (SISO) formulation inherently neglects multi-axis coupling effects and relies on heuristic tuning procedures, rendering them susceptible to parametric uncertainties and exogenous disturbances \cite{kasnakouglu2016investigation, kada2011robust}. Gain scheduling partially compensates for nonlinearities by adapting controller parameters across the operational envelope, yet this approach provides only incremental improvements in robustness while necessitating extensive empirical calibration \cite{melo2022fuzzy, abbas2024survey}.

Linear Quadratic Regulators (LQR) provide a mathematically optimal control framework that minimizes a quadratic cost function, balancing state regulation performance against control effort. Despite their theoretical optimality, LQR implementations suffer from two critical limitations: they require full state feedback, necessitating state estimation in practical applications, and they lack inherent robustness guarantees against parametric uncertainties. When system dynamics deviate from the nominal model -- a common occurrence in aerial vehicles due to aerodynamic effects and mass distribution changes -- LQR performance deteriorates significantly. Additionally, the state estimation typically relies on Kalman filtering, which achieves optimality only under Gaussian noise assumptions. Furthermore, LQR controllers commonly exhibit longer response times than their PID counterparts, despite their theoretical optimality properties \cite{elkhatem2022robust}.

Nonlinear control methodologies have been proposed to address the limitations above. Backstepping techniques, founded on recursive Lyapunov stability theory, offer enhanced robustness to matched uncertainties. However, their implementation yields controllers of elevated structural complexity with substantial computational overhead and parameter sensitivity, thus imposing practical constraints on real-time applications \cite{abbas2024survey}. Feedback linearization approaches transform nonlinear dynamics into equivalent linear systems via nonlinear state transformations and control laws. Yet, these methods exhibit pronounced sensitivity to modeling errors and measurement noise, consequently compromising robustness margins \cite{abbas2024survey}. Adaptive control frameworks facilitate online parameter estimation and controller reconfiguration to mitigate parametric uncertainties.
Nevertheless, several challenges persist, including parameter convergence rates under time-varying conditions, susceptibility to measurement noise, and analytical complexity in establishing uniform stability guarantees. While these nonlinear methodologies demonstrate theoretical superiority over linear control architectures under nominal conditions, they frequently encounter implementation barriers on resource-constrained UAV platforms and lack rigorous performance guarantees under structured uncertainties and bounded disturbances. 

These limitations have motivated the development of robust control frameworks, particularly $\mathcal{H}_\infty$ control \cite{zhou1998essentials}, which provides rigorous stability guarantees under bounded parameter variations by representing system uncertainties as norm-bounded operators and minimizing worst-case induced norms without requiring precise disturbance characterization. These methodologies have gained significant traction across aerospace applications, with recent literature demonstrating their effectiveness for tailsitter UAVs under severe turbulence \cite{kumar2024h}, helicopter control with certified performance bounds \cite{gadewadikar2008structured}, and resource-constrained implementations that substantially outperform traditional PID architectures despite computational limitations \cite{bautista2008implementation}.

A key advantage of robust control frameworks lies in their compatibility with Linear Parameter-Varying (LPV) modeling, which enables exact representation of nonlinear dynamics without local linearization or approximation \cite{kumar2025sparse}. The rotational dynamics of multi-rotors contain trigonometric and rational expressions that can be modeled precisely using static nonlinear operators, which are then treated as structured, bounded uncertainty blocks within the Linear Fractional Transformation (LFT) formalism. This representation maintains the full fidelity of the nonlinear dynamics while enabling systematic controller synthesis via the $\mathcal{H}_\infty$ technique.

The LPV-LFT approach effectively bridges the gap between high-fidelity dynamic modeling and tractable robust control design by encapsulating nonlinearities as parameter-dependent terms around a linear time-invariant core. This framework embeds parameter variations directly into the synthesis process rather than treating them as afterthoughts, offering formal robustness guarantees against structured uncertainties. The resulting controllers maintain stability and performance across the entire operating envelope without requiring gain scheduling or extensive empirical tuning, making them particularly suitable for autonomous systems operating in uncertain environments.

The effectiveness of LFT-based modeling has been demonstrated across aerospace applications, including NASA's stability margin assessments \cite{shin2008robustness} and ESA's spacecraft attitude control \cite{di2010integrated}. LPV approaches extend beyond gain scheduling by embedding parameter variations into control design, offering formal robustness guarantees \cite{shamma1990analysis, shamma2012overview}. While synthesis methods typically use linear matrix inequalities \cite{el2000advances, helton2007linear}, challenges remain in representing nonlinear dependencies, often requiring probabilistic solutions \cite{tempo2013randomized, fujisaki2003probabilistic}. Nevertheless, LPV-LFT approaches have achieved success in applications from reconfigurable flight control to polynomially parameterized stability analysis \cite{ganguli2002reconfigurable, balas2002linear, gilbert2010polynomial, marcos2009lpv}, motivating this work's robust control framework for aerospace systems.

\subsection{New Contributions}

This paper addresses significant challenges in robust multi-rotor stabilization by advancing the application of structured uncertainty frameworks to nonlinear multi-rotor flight dynamics. The contributions of this work are threefold:

\begin{enumerate}
  \item We establish a systematic framework for representing multi-rotor nonlinear dynamics within the LFT formalism, enabling rigorous treatment of trigonometric nonlinearities and parameter-dependent terms as structured uncertainties.
  
  \item We develop and validate robust control synthesis techniques that utilize $\mathcal{H}_\infty$ methodology for attitude stabilization under significant external disturbances, utilizing only gyroscope measurements with explicit characterization of state-dependent sensor noise.
  
  \item We provide a comprehensive comparative analysis between classical PID and robust control approaches, quantifying performance improvements regarding disturbance rejection capabilities and control effort optimization under wind turbulence conditions representative of practical operating environments.
\end{enumerate}

This work studies the nonlinear flight dynamics of a multirotor aerial vehicle within a structured uncertainty framework. The equations of motion considered in this paper correspond to the standard nonlinear rigid-body dynamics commonly used for generic UAV platforms. The formulation is not specific to any particular vehicle configuration and it represents a general multirotor model. The numerical parameter values used in simulations correspond to a representative vehicle for evaluation. Building on this modeling framework, this work advances the application of structured uncertainty frameworks in UAV control by integrating LPV-LFT modeling with robust synthesis techniques. Unlike approaches that rely on linearization approximations, our methodology precisely captures system nonlinearities while incorporating realistic sensor limitations. The proposed framework addresses fundamental challenges in existing multi-rotor controllers by providing formal robustness guarantees against parametric uncertainties and exogenous disturbances, thereby bridging theoretical control design with practical implementation constraints.

\section{LPV Modeling of Multi-Rotor Systems}
\subsection{Dynamics}
\label{LFT Modeling of the Dynamics}
We consider the classical multi-rotor UAV dynamics derived from the Newton-Euler equations. Since this work focuses on robust stabilization, the analysis is restricted to the rotational dynamics of the vehicle, while the translational motion is omitted. The nonlinear equations of motion for the rotational dynamics of a multi-rotor UAV are given in \cite{kumar2024h} with ZYX Euler sequence:
\begin{equation}
\label{EOM1}
\begin{aligned}
   \begin{bmatrix} \dot{\phi}\\ \dot{\theta}\\\dot{\psi} \end{bmatrix} = 
   \begin{bmatrix}
        1 & \sin{\phi}\tan{\theta}  & \cos{\phi}\tan{\theta} \\ 
        0 & \cos{\phi} & -\sin{\phi} \\
        0 & \sin{\phi}\sec{\theta} & \cos{\phi}\sec{\theta}
   \end{bmatrix}
   \begin{bmatrix} p\\q\\ r \end{bmatrix}
\end{aligned},
\end{equation}

\begin{equation}
\label{EOM2}
\begin{aligned}
   \begin{bmatrix} \dot{p}\\\dot{q}\\\dot{r} \end{bmatrix}=
   \begin{bmatrix}
       \frac{I_y-I_z}{I_x}rq + \frac{L}{I_x}  \\
       \frac{I_z-I_x}{I_y}pr + \frac{M}{I_y}  \\
       \frac{I_x-I_y}{I_z}pq + \frac{L}{I_z}         
   \end{bmatrix},
\end{aligned}
\end{equation}
where $\begin{bmatrix} \phi & \theta & \psi \end{bmatrix}^\top$ are Euler angles, $\begin{bmatrix} p & q & r \end{bmatrix}^\top$ are angular rates in the body frame, $\texttt{blkdiag}(I_x,I_y,I_z)$ is the inertia matrix, and $\begin{bmatrix} L & M & N \end{bmatrix}^\top$ are the roll, pitch, and yaw moments generated by the propellers.

The nonlinearities in Eqn. \eqref{EOM1}-\eqref{EOM2} arise from trigonometric and state-dependent product terms. To streamline the robust controller synthesis, these dynamics are expressed in LPV form as

\begin{multline}
\label{LPV}
      \begin{bmatrix}\dot{\phi}\\ \dot{\theta}\\\dot{\psi}\\ \dot{p}\\\dot{q}\\\dot{r} \end{bmatrix}
   = J 
   \begin{bmatrix}
       0 & 0 & 0 & 1 & \frac{\rho_1\rho_3}{\rho_4} & \frac{\rho_2\rho_3}{\rho_4}  \\
       0 & 0 & 0 & 0 & \rho_2 & -\rho_1  \\
       0 & 0 & 0 & 0 & \frac{\rho_1}{\rho_4} & \frac{\rho_2}{\rho_4}\\
       0 & 0 & 0 & 0 & \frac{\rho_7}{2} & \frac{\rho_6}{2}\\
       0 & 0 & 0 & \frac{\rho_7}{2} & 0 & \frac{\rho_7}{2}\\
       0 & 0 & 0 & \frac{\rho_6}{2} & \frac{\rho_5}{2}  & 0\\
   \end{bmatrix}\begin{bmatrix} \phi \\ \theta \\ \psi \\ p \\ q \\ r \end{bmatrix}\\  + \begin{bmatrix}
       0 & 0 & 0 \\
       0 & 0 & 0 \\
       0 & 0 & 0 \\
       \frac{1}{I_x} & 0 & 0 \\
       0 & \frac{1}{I_y} & 0 \\
       0 & 0 & \frac{1}{I_z} \\
   \end{bmatrix} \begin{bmatrix} L \\M\\N \end{bmatrix},
\end{multline}
where the parameters are defined as 
\begin{equation}
\label{params}
\begin{aligned}
    \begin{bmatrix}\rho_1\\ \rho_2\\ \rho_3\\ \rho_4 \end{bmatrix} 
      &= \begin{bmatrix} \sin{\phi}\\ \cos{\phi} \\ \sin{\theta} \\\cos{\theta} \end{bmatrix},
    \begin{bmatrix}\rho_5\\ \rho_6\\ \rho_7 \end{bmatrix} 
      &= \begin{bmatrix} p \\ q \\ r \end{bmatrix},
\end{aligned}
\end{equation}
and $J = \texttt{blkdiag}(I_{3\times3}, \tfrac{I_y-I_z}{I_x}, \tfrac{I_z-I_x}{I_y}, \tfrac{I_x-I_y}{I_z})$ is the inertia multiplier matrix. The formulation is linear in the states but nonlinear in the parameter set $\rho$. The parameters are bounded by trigonometric identities and actuator constraints, i.e., $\rho \in [\rho_{\min}, \rho_{\max}]$.

\subsection{Measurement Model and Sensor Noise}
In practice, the onboard IMU sensors provide only angle rate measurements. Therefore, we consider only the gyroscopic sensor measurements in the robust controller synthesis. Accordingly, the measurement model is

\begin{equation}
\label{measurement}
\begin{aligned}
    y_m
   = 
   \begin{bmatrix}
       0 & 0 & 0 & 1 & 0 & 0 \\
       0 & 0 & 0 & 0 & 1 & 0 \\
       0 & 0 & 0 & 0 & 0 & 1 \\
       \end{bmatrix}  \begin{bmatrix} \phi \\ \theta \\ \psi \\ p \\ q \\ r \end{bmatrix} + \nu,
\end{aligned}
\end{equation}
where $\nu$ represents the noise of the gyroscopic sensor. 

While the measurement model is linear, practical IMU noise exhibits frequency-dependent characteristics due to internal sensor dynamics and filtering components. This becomes particularly problematic during aggressive maneuvers, where signal and noise frequency content overlap substantially. To accurately capture these effects within our robust control framework, sensor noise is shaped by frequency-dependent weights, which characterize the spectral distribution of measurement noise.


\begin{equation}
\label{noise}
    \nu = \texttt{blkdiag}(W_\phi(j\omega),W_\theta(j\omega),W_\psi(j\omega))\Bar{\nu}
\end{equation}
where $\Bar{\nu}$ is unit-norm exogenous noise, and $W_{\phi,\theta,\psi}(j\omega)$ are frequency-dependent weighting functions, typically modeled using the sensor noise characteristics provided by the manufacturer.

\subsection{Robust Control Problem as an LFT Interconnection}
The multi-rotor dynamics in Eqn. \eqref{LPV}, together with the measurement model in Eqn. \eqref{measurement}-\eqref{noise}, can be expressed in the LPV state-space form

\begin{subequations}
\label{sys}
    \begin{align}
        \dot{x} &= A(\rho) x + B_u(\rho)u + B_ww,\\
        y_m &= C_y(\rho)x + D_u(\rho) + D_ww,\\
        z &= C_zx + D_z u,
    \end{align}
\end{subequations}
where $x$ is the state vector, $y_m$ is the measured output, and $w$ is the exogenous input that includes sensor noise and disturbance. $z$ represents the regulated outputs that need to be minimized, which in our case is the attitude error and actuator efforts. The vector $\rho$ represents the LPV parameters given in \eqref{params}. The system matrices are nonlinear functions of $\rho$.

\begin{figure}[ht]
\centering
  \includegraphics[scale= .6]{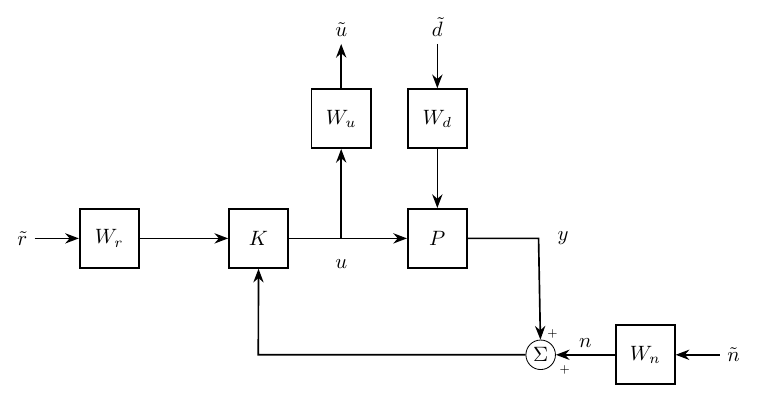}
  \caption{System interconnection for designing and implementing proposed controllers. }
  \label{fig:interconnections}
\end{figure}

The control objective is to design an output-feedback controller $K$ that regulates $z$ based on the available measurement $y_m$, while ensuring robust stability and performance against parameter variations $\rho$ and exogenous disturbances $w$. This is achieved by embedding the system into the generalized interconnection shown in Fig.~\ref{fig:interconnections}, where weighting functions $W_{r,d,u,n}$ map the physical design specifications into the $\mathcal{H_\infty}$ framework. 

These weights shape the frequency response and capture actuator bandwidth limits, noise spectra, and disturbance rejection requirements that are not directly visible in the raw signals. Moreover, the exogenous inputs are normalized as unit-norm bounded, ensuring that the $\mathcal{H}_\infty$ optimization problems remain well-posed, with performance guarantees scaling proportionally with disturbance magnitudes.

Formally, the objective is to define a tunable state space model of the controller to minimize
\begin{equation}
\| z\|_2 = \| T_{w \to z}(s)\|_{\mathcal{H}_\infty},
\end{equation}
where $T_{w \to z}$ is the closed-loop transfer matrix from $w$ to $z$. The $\mathcal{H}_\infty$ objective ensures attenuation of worst-case disturbances, while accounting for structured uncertainties in $\rho$.

\begin{figure}[ht]
 \centering
   \includegraphics[scale= .6]{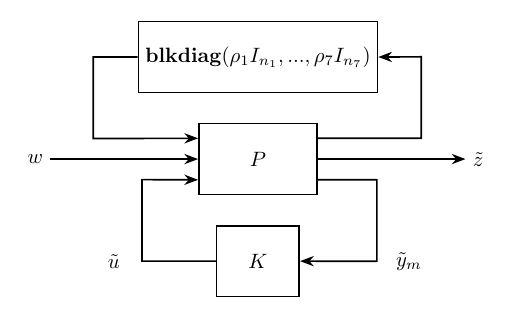}
   \caption{LFT interconnection for designing robust controllers.}
   \label{fig:lft}
 \end{figure}

To synthesize the controller, the nonlinear parameter dependence in Eqn. \eqref{sys} is expressed as an LFT interconnection (Fig.~\ref{fig:lft}) by treating the $\rho$ terms as structured uncertainty blocks. In MATLAB’s Robust Control Toolbox, this is implemented by declaring $\rho$ as \texttt{ureal} objects, automatically constructing the uncertainty blocks for LFT-based analysis. The controller $K$ is then synthesized using \texttt{hinfstruct(...)} or \texttt{hinfsyn(...)}.

A key technical challenge in the LFT formulation is that the nominal plant contains poles at the origin, a characteristic feature of multi-rotor dynamics due to their free-body rotational modes. These poles violate the well-posedness conditions required for standard $\mathcal{H}_\infty$ optimization. To address this issue, we implement stability augmentation through minimal viscous aerodynamic/motor damping terms that shift the poles slightly into the left-half plane. This modification has a negligible effect on the physical fidelity of the model but renders the optimization problem mathematically tractable.

\section{Simulation Results}

\subsection{Design Parameters}
For robust control synthesis, the physical parameters of the multi-rotor are summarized in \autoref{tab:physical params}, and the parameter bounds $\rho$ in Eqn.~\eqref{params} are determined by the operational envelope and actuator constraints, as listed in \autoref{tab:bounds}.

\renewcommand{\arraystretch}{1.5}
\begin{table}[ht]
\caption{Physical parameters of the multi-rotor}
\centering
\begin{tabular}{c||c|c|c|c}
  \textbf{Parameter} & mass & $I_x$ & $I_y$ & $I_z$ \\ \hline
  \textbf{Value} & 10 kg & 0.25 kg$\cdot$m$^2$ & 0.2 kg$\cdot$m$^2$ & 0.1 kg$\cdot$m$^2$
  \end{tabular}
\label{tab:physical params}
\end{table}

\begin{table}[ht]
\caption{Parameter bounds}
\centering
\begin{tabular}{c||c|c|c|c|c|c|c}
  \textbf{Parameter} & $\rho_1$ & $\rho_2$ & $\rho_3$ & $\rho_4$ & $\rho_5$ & $\rho_6$ & $\rho_7$ \\ \hline
  $\mathbf{\rho_{min}}$ & -1 & 0 & -1 & 0 & -1.5 r/s & -1.5 r/s & -1.5 r/s  \\ \hline
  $\mathbf{\rho_{max}}$ & 1 & 1 & 1 & 1 & 1.5 r/s & 1.5 r/s & 1.5 r/s
  \end{tabular}
\label{tab:bounds}
\end{table}

The plant is augmented by incorporating actuator dynamics and weighting functions to map the physical design requirements into the $\mathcal{H}_\infty$ framework. The actuators are modeled as first-order systems, and constant weighting functions are employed. This choice simplifies the synthesis problem by avoiding additional dynamics in the augmented plant while capturing the relative importance of different performance channels. The constant weights act as scaling factors that normalize these channels, ensuring each contributes appropriately to the overall performance index. This augmentation yields the standard interconnection structure shown in Fig. \ref{fig:interconnections}, ensuring that performance constraints are explicitly represented in the closed-loop problem.

The controller is synthesized using MATLAB’s \texttt{hinfsyn(...)}, which formulates the uncertain blocks as an upper LFT and minimizes the $\mathcal{H}_\infty$ norm of the augmented system. The resulting controller has nine states, corresponding to the three disturbance channels, three sensor noise channels, and three actuator effort channels that are jointly minimized in the performance objective. In contrast, the cascaded SISO PID controller uses two states per axis, for a total of six states.

The closed-loop performance achieved by the $\mathcal{H}_\infty$ controller is quantified by the norm $\gamma_0 = 0.25$. This indicates that the worst-case amplification from disturbances and sensor noise to the regulated outputs is bounded by $\gamma_0$ in the induced $\mathcal{L}_2$ sense, providing strong robustness and disturbance attenuation guarantees.

\subsection{Simulation Setup}
Multi-rotor missions are particularly sensitive during takeoff, hover, and landing, as these phases are heavily affected by unmodeled dynamics such as ground effects and wind gusts. This makes crucial tasks, including VTOL and payload delivery in a constrained environment, challenging. To emulate these conditions, we consider a simulation in which a multi-rotor UAV is tasked with hovering at an altitude of 10 m in the presence of strong winds. 

The exogenous input vector is considered as
\begin{equation}
    w(t) = [ \tau_\phi(t) \quad \tau_\theta(t) \quad \tau_\psi(t) \quad n_p \quad n_q \quad n_r]^\top
\end{equation}
where $\tau_{\phi,\theta,\psi}(t)$ denote disturbance torques induced by wind gusts along the respective axes, and $n_{p,q,r}(t)$ represent sensor noise in gyroscopic measurements.

For validation, a nonlinear Simulink model of the complete system is employed. In Simulink, a three-axis gyroscope block with an SNR of 35 dB is used, representative of commercially available MEMS IMUs. Wind disturbances are modeled using the Dryden turbulence model, consistent with military specifications to simulate gust velocities up to 15 m/s, corresponding to worst-case operational conditions. 

For baseline comparison, a cascaded SISO PID controller is implemented for each axis, owing to its popularity. The gains are tuned using MATLAB’s autotune feature to ensure best-case PID performance. This autotuner works by performing a frequency-response estimation experiment to identify the system dynamics, and then calculating PID gains based on desired performance metrics. The  $\mathcal{H}_\infty$ controller is evaluated against this benchmark.

\subsection{Results and Discussions}
\begin{figure}[ht]
\centering
  \includegraphics[scale= 0.45]{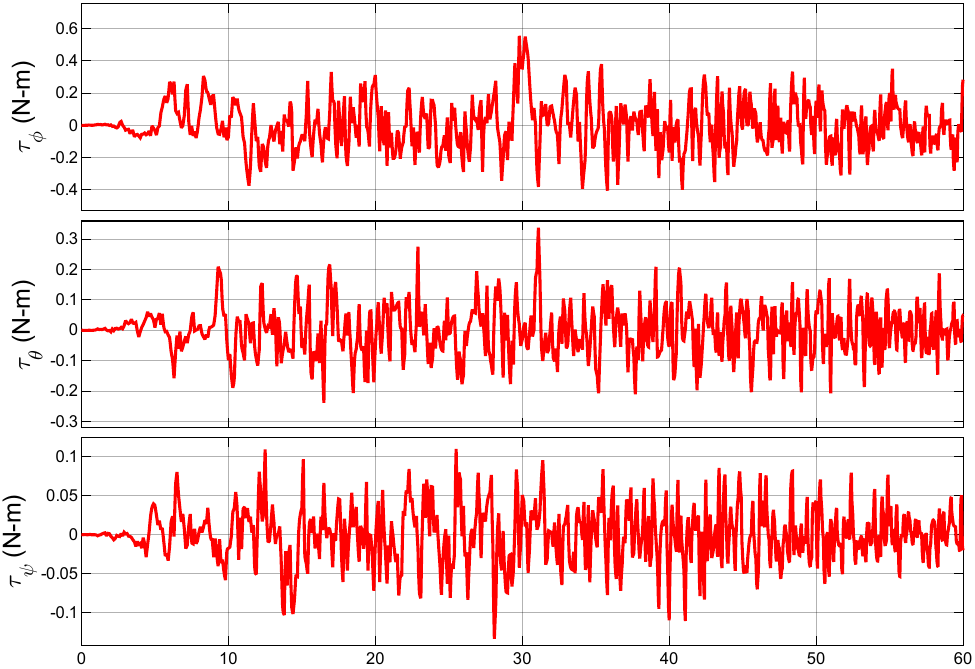}
  \caption{Disturbance moments generated by wind gust and turbulence}
  \label{fig:dist_moment}
\end{figure}

The disturbance moments generated by the Dryden turbulence model are shown in Fig. \ref{fig:dist_moment}. The Dryden turbulence model produces disturbance moments up to $0.65$ N-m in roll and pitch, corresponding to moderate-to-severe gust conditions. This replicates practical scenarios such as payload delivery in urban areas subject to wind shear or missions in hilly terrain with strong localized turbulence. 


\begin{figure}[ht]
\centering
  \includegraphics[scale= 0.5]{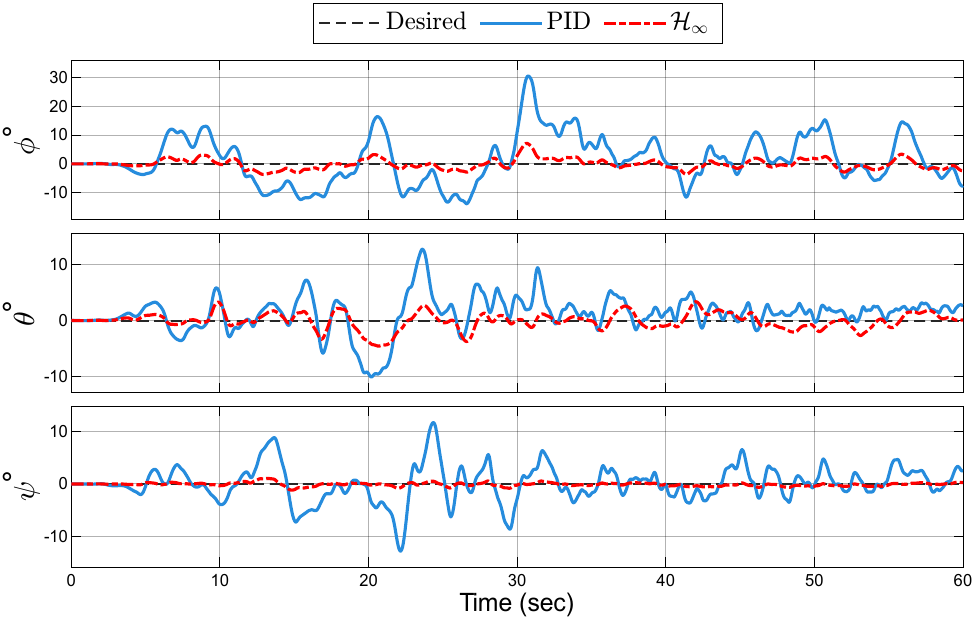}
  \caption{Euler angles (attitude)  of the multi-rotor}
  \label{fig:attitude}
\end{figure}

\begin{figure}[ht]
\centering
  \includegraphics[scale= 0.5]{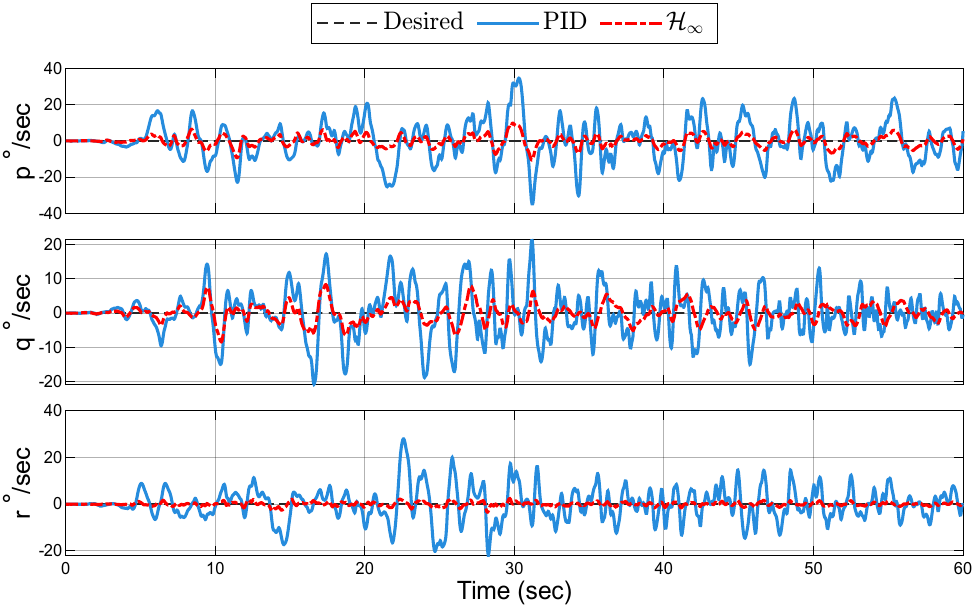}
  \caption{Rotational rates of the multi-rotor}
  \label{fig:pqr}
\end{figure}

The closed-loop attitude trajectories under PID and $\mathcal{H}_\infty$ controllers are presented in Fig. \ref{fig:attitude}-\ref{fig:pqr}. Since it is a regulation problem, the controllers reject disturbances and drive the states to zero using only gyroscope measurements. The results demonstrate that the robust controller achieve significantly superior performance compared to the PID controller, showing improved resilience against disturbances, uncertainties, and measurement noise. In all cases, the errors remain bounded, validating the robustness of the proposed controllers against dynamic and measurement uncertainties.



The Euler angle deviations' peak and root mean square error (RMSE) values are summarized in \autoref{tab:errors}. The PID controller exhibits large deviations, with a peak error exceeding $30^\circ$, whereas the $\mathcal{H}_\infty$ controller reduces the error by nearly an order of magnitude. 

\begin{table}[ht]
\caption{Errors in multi-rotor attitude using different controllers}
\centering
\begin{tabular}{c||c|c}
   & $\mathbf{PID}$ & $\mathbf{\mathcal{H}_\infty}$ \\ \hline
   \textbf{Peak magnitude} & 30.54$^\circ$ &  7.12$^\circ$ \\ \hline
   \textbf{RMSE} & 5.67$^\circ$ & 1.35$^\circ$ 
  \end{tabular}
\label{tab:errors}
\end{table}

\begin{figure}[ht]
\centering
  \includegraphics[scale= 0.5]{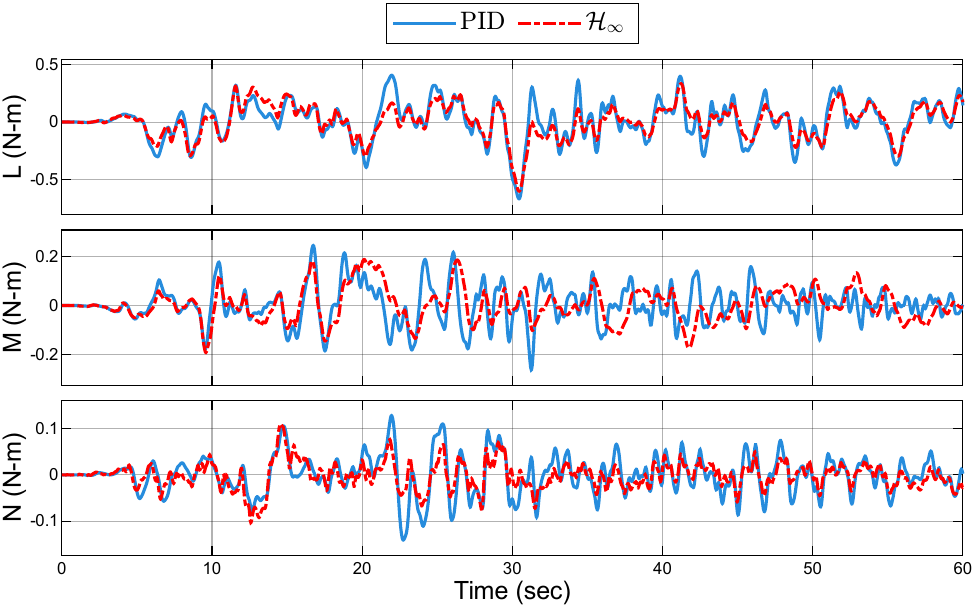}
  \caption{Control Inputs}
  \label{fig:control_ip}
\end{figure}

The control inputs required to achieve the desired states are shown in Fig. \ref{fig:control_ip}. The $\mathcal{H}_\infty$ controller achieves lower tracking errors without exerting aggressive or high-magnitude actuator efforts. Additionally, the actuator model in the simulation ensures that the controller does not exceed actuator bandwidths. Therefore, the $\mathcal{H}_\infty$ controller exhibits overall improved performance relative to the PID controller as it regulates the states effectively with less actuator exertion. 


We also verified that the LPV parameters $\rho(t)$ remain within the assumed bounds under closed-loop operation. Since roll and pitch are constrained to $\phi,\theta \in [\frac{-\pi}{2},\frac{\pi}{2}]$, the trigonometric scheduling terms (e.g., $\sin{(.)}, \cos{(.)}$) are naturally bounded in $[-1,1]$. The states $p,q,r$ (corresponding to $\operatorname{\rho_5 - \rho_7}$) remain within the prescribed limits ($\pm 1.5\pi$ rad) throughout the simulations, as confirmed by the state trajectories shown in Fig. \ref{fig:pqr}. No violations of the $\rho$ bounds were observed for the $\mathcal{H}_\infty$ controller.

\section{Conclusions}
This paper presents a robust multi-rotor UAV attitude stabilization control framework that systematically captures nonlinear dynamics within the Linear Fractional Transformation (LFT) framework. The proposed $\mathcal{H}_\infty$ methodology preserves full dynamic fidelity in the controller synthesis by treating trigonometric nonlinearities and parameter variations as structured uncertainties rather than linearization approximations. The approach utilizes only gyroscope measurements and provides stability guarantees.

Simulation results under realistic conditions demonstrate substantial performance improvements over classical PID control, with significant reductions in peak attitude errors and RMSE values under severe wind disturbances generated by the Dryden turbulence model. These improvements are achieved while maintaining lower actuator effort and operating with realistic gyroscope noise representative of commercial MEMS IMUs. The validation confirms the practical applicability of $\mathcal{H}_\infty$  theory for multi-rotor attitude control in challenging missions such as urban navigation, VTOL operations, and payload delivery in gusty environments.

\bibliographystyle{IEEEtran}
\bibliography{refs}

\end{document}